\documentclass[12pt]{article}
\usepackage{geometry}
\usepackage{amsmath}
\usepackage{amsfonts}          
\usepackage{amssymb}
\usepackage{xspace}
\usepackage{graphicx}

\def\appendix#1{
  \addtocounter{section}{1}
  \setcounter{equation}{0}
  \renewcommand{\thesection}{\Alph{section}}
 \section*{Appendix \thesection\protect\indent \parbox[t]{11.715cm} {#1}}
  \addcontentsline{toc}{section}{Appendix \thesection\ \ \ #1}
  }

\renewcommand{\thefootnote}{\fnsymbol{footnote}}

\numberwithin{equation}{section}

\newcommand{\be}{\begin{equation}}
\newcommand{\ee}{\end{equation}}
\newcommand{\ba}{\begin{aligned}}
\newcommand{\ea}{\end{aligned}}

\newcommand{\ie}{{\it i.e.}}
\newcommand{\eg}{{\it e.g.}}
\newcommand{\cf}{{\it cf.}}

\newcommand{\sltwo}{\mathfrak{sl}(2)}
\newcommand{\sutwo}{\mathfrak{su}(2)}

\makeatletter
\def\sla@#1#2#3#4#5{{%
  \setbox\z@\hbox{$\m@th#4#5$}%
  \setbox\tw@\hbox{$\m@th#4#1$}%
  \dimen4\wd\ifdim\wd\z@<\wd\tw@\tw@\else\z@\fi
  \dimen@\ht\tw@
  \advance\dimen@-\dp\tw@
  \advance\dimen@-\ht\z@
  \advance\dimen@\dp\z@
  \divide\dimen@\tw@
  \advance\dimen@-#3\ht\tw@
  \advance\dimen@-#3\dp\tw@
  \dimen@ii#2\wd\z@
  \raise-\dimen@\hbox to\dimen4{%
    \hss\kern\dimen@ii\box\tw@\kern-\dimen@ii\hss}%
  \llap{\hbox to\dimen4{\hss\box\z@\hss}}}}
\def\slashed#1{%
  \expandafter\ifx\csname sla@\string#1\endcsname\relax
    {\mathpalette{\sla@/00}{#1}}%
  \else
    \csname sla@\string#1\endcsname
  \fi}
\makeatother

\begin{document}

%%%%%%%%%%%%%%%%%%%%%%%%%%%%%%%%%%%%%%%%%%%%%%%%%%%%%%%%%%%%%%%%%%%%%%%%

\thispagestyle{empty}
\begin{flushright}\footnotesize
\texttt{hep-th/0602214}\\
\texttt{DESY-06-018}\\
\texttt{ZMP-HH/06-02}\\
\vspace{2.1cm}
\end{flushright}

\renewcommand{\thefootnote}{\fnsymbol{footnote}}
\setcounter{footnote}{0}

\begin{center}
{\Large\textbf{\mathversion{bold} 
Exact expressions for quantum corrections to \\
spinning strings 
}\par}

\vspace{2.1cm}

\textrm{Sakura Sch\"afer-Nameki} 

\vspace{1cm}

\textit{II. Institut f\"ur Theoretische Physik der Universit\"at Hamburg\\
Luruper Chaussee 149, 22761 Hamburg, Germany} \\

\vspace{1.5mm}
\textit{and}
\vspace{1.5mm}

\textit{Zentrum f\"ur Mathematische Physik, Universit\"at Hamburg\\
Bundesstrasse 55, 20146 Hamburg, Germany} \vspace{3mm}

\vspace{1.5mm}
\texttt{sakura.schafer-nameki at desy.de}

\vspace{3mm}

%%%%%%%%

\par\vspace{1cm}

\textbf{Abstract}

\vspace{5mm}

\end{center}
The one-loop worldsheet 
quantum corrections to the energy of spinning strings on $ \mathbb{R} \times S^3$ within  
$AdS_5 \times S^5$ are reexamined. 
The explicit expansion in the effective 't Hooft coupling $\lambda'= \lambda/J^2$ is rigorously derived. The expansion contains both analytic and non-analytic terms in $\lambda'$, as well as exponential corrections. Furthermore, we pin down the origin of the terms that are not captured by the quantum string Bethe ansatz, which only produces analytic terms in $\lambda'$. It is shown that the analytic terms
arise from string fluctuations within the $S^3$, whereas 
the non-analytic and exponential terms, 
which are not captured by the Bethe ansatz, originate from the fluctuations in all directions within the supersymmetric sigma model 
on $AdS_5 \times S^5$. We also comment on the case of spinning string in $AdS_3 \times S^1$.

\vspace*{\fill}

\newpage
\setcounter{page}{1}
\renewcommand{\thefootnote}{\arabic{footnote}}
\setcounter{footnote}{0}

%%%%%%%%%%%%%%%%%%%%%%%%%%%%%%%%%%%%%%%%%%%%%%%%%%%%%%%%%%%%%%%%%%%%%%%

%\tableofcontents

%%%%%%%%%%%%%%%%%%%%%%%%%%%%%%%%%%%%%%%%%%%%%%%%%%%%%%%%%%%%%%%%%%%%%%%%

\section{Introduction and Summary}

The world-sheet one-loop corrections to the energy of spinning strings in $AdS_5 \times S^5$ has been the subject of vivid discussions. 
A better understanding of these quantum string corrections would not only
elucidate various aspects of the AdS/CFT correspondence between string theory 
on $AdS_5 \times S^5$ and $d=4$, $\mathcal{N}=4$ $SU(N_c)$ SYM theory, but would moreover 
provide valuable insight into the structure of quantum strings on curved, 
flux-supported backgrounds, which so far are not amenable to standard quantization techniques.
 
A bold and possibly very powerful conjecture was put forward, packaging the 
complete quantum string spectrum on $AdS_5\times S^5$ into a Bethe ansatz \cite{Arutyunov:2004vx, Staudacher:2004tk, Beisert:2005fw}. This proposal was partly inspired by the 
Bethe ansatz description of anomalous dimensions of gauge-invariant operators in SYM 
\cite{Minahan:2002ve, Beisert:2003yb, Serban:2004jf, Beisert:2004hm, Staudacher:2004tk, Beisert:2005fw, Rej:2005qt}, 
and likewise the existence of a Bethe-ansatz-like structure for the classical string on $AdS_5 \times S^5$ 
\cite{ Kazakov:2004qf, Zarembo:2004hp, Beisert:2004ag, Schafer-Nameki:2004ik, 
Beisert:2005bm, Alday:2005gi}. Needless to say, testing this Bethe ansatz is of utmost importance. 

A particularly restrictive constraint that has to be met by the Bethe ansatz are 
the world-sheet corrections to the 
Frolov-Tseytlin solutions that can be computed semi-classically
\cite{Frolov:2003tu, FrolovTseytlinI,Frolov:2003qc,Frolov:2004bh, Park:2005ji}.  
The present status of these investigations is that the string Bethe ans\"atze capture these semi-classical results only partly \cite{Beisert:2005mq, Hernandez:2005nf, Schafer-Nameki:2005tn, BT, Schafer-Nameki:2005is}.
The subject of this letter is to pinpoint the problem which is causing the disagreement.

The one-loop energy shift, \ie, the $O(\alpha')$ corrections to the string energy, 
has been discussed at leading order in the 't Hooft coupling $\lambda' =\lambda/J^2$ in \cite{Beisert:2005mq, Hernandez:2005nf}. In particular, the analysis of \cite{Beisert:2005mq} showed, that these corrections could be computed from the Landau-Lifschitz model, which arises from the $S^3$ sector of the sigma model.
In \cite{Schafer-Nameki:2005tn} a thorough investigation of the comparison between Bethe ansatz and semi-classical strings was undertaken, concluding, that under the assumption that a certain zeta-function regularization  is applicable, there is agreement at least up to order ${\lambda'}^3$ and furthermore the string energy has an analytic expansion in $\lambda'$. However, the semi-classical strings and Bethe ansatz expressions were also shown to disagree when expanded for large winding numbers. 
This was the first indication that the Bethe ansatz may not entirely reproduce the semi-classical result.

In addition to the large winding number discrepancy, one can convince oneself of 
the limitations of zeta-function regularization, which can be pinned down
already on the level of relatively simple sums \cite{Schafer-Nameki:2005is}. 
Applying an integral approximation to the one-loop energy shift in the $\sutwo$ sector, \ie, spinning strings on $S^3 \times \mathbb{R}$, it was argued that the $\lambda'$-expansion contains not only the analytic terms that arise from zeta-function regularization, but also contains non-analytic terms of order ${\lambda'}^{(2n+1)/2}$ \cite{BT, Schafer-Nameki:2005is, Minahan:2005qj} and possibly exponential corrections of order $e^{-\lambda'}$ \cite{Schafer-Nameki:2005is}. Furthermore, neither of these are captured by the string Bethe ansatz. In \cite{BT} a proposal was put forward, which corrects the Bethe ansatz in order to incorporate the non-analytic terms. 

The purpose of this letter is to derive the exact expression for the coefficients in
 the $\lambda'$-expansion of the one-loop worldsheet correction in the $\sutwo$-case as computed 
from a semi-classical analysis in \cite{Frolov:2003tu, Frolov:2004bh}. 

Before summarizing our findings, let us briefly recall the structure of the one-loop energy shift. 
Consider the classical spinning string solution to the supersymmetric sigma-model on $AdS_5 \times S^5$, which is 
supported on $S^3 \times \mathbb{R}$ and carries $S^3$ angular momentum $J$. 
Classically, the system is fully described by the fields on $S^3 \times \mathbb{R}$, and the remaining directions in the sigma-model decouple. 
This is however no longer the case for the quantum corrections, which are a sum 
over characteristic frequencies with contributions from all
fields within the supersymmetric $AdS_5 \times S^5$ sigma-model (in the present case 
there are two $S^3$ fluctuation modes, in addition to the transverse six bosonic and eight fermionic fluctuations). 
In particular, the direct quantization 
of the classical reduced system need not lead to the correct quantum string spectrum, 
as was \eg\ observed in \cite{Arutyunov:2005hd}. 
Note that this is quite different to the dual gauge theory, where \eg, the $\sutwo$ subsector remains closed to all loops, see also \cite{Minahan:2005jq}.

Expanding the sum over frequencies at $O(\alpha')$ in a series in the effective 't Hooft coupling $\lambda'$, we find the following: 
\begin{itemize}
\item Analytic terms ${\lambda'}^n$: $S^3$ modes
\item Non-analytic terms ${\lambda'}^{(2n+1)/2}$: $S^3$, transverse bosonic and fermionic modes
\item Exponential terms $e^{-\lambda'}$: $S^3$, transverse bosonic and fermionic modes.
\end{itemize}
This in particular confirm the leading order in $\lambda'$ result of \cite{Beisert:2005mq}, where it was shown that the 
analytic terms can be reproduced from the Landau-Lifschitz model, which only sees the $S^3$-part of the fluctuations. Furthermore this is in 
agreement with the analytic terms arising from zeta-function regularization in
\cite{Schafer-Nameki:2005tn, Schafer-Nameki:2005is}. 
The non-analytic terms confirm the ones in \cite{BT, Minahan:2005qj}, 
where they were constructed by means of an Euler-Maclaurin type integral approximation to the sum over fluctuation frequencies. 
The procedure which we apply systematically incorporate all these results, and furthermore shows the existence of the exponentially suppressed terms. 

Some comments are in order:
firstly, one should keep in mind, that in the $\sutwo$ sector, 
which we study here in detail, the solution is not stable for 
arbitrary choices of the parameter $k$. In particular, 
the fluctuation frequencies become complex for $2 k > 1$. 
One therefore has to analytically continue the expression for the energy in $k$. 
This instability can also be seen from the Bethe ansatz, 
as was discussed in \cite{Beisert:2005mq}. Therefore, any discussion of
the $\sutwo$ sector needs to be taken with a grain of salt. 

Keeping this in mind, one can nevertheless 
investigate the comparison to the Bethe ansatz. 
The Bethe ansatz of \cite{Arutyunov:2004vx} captures precisely the analytic terms, however misses out the non-analytic  and exponential corrections. 
Put differently, our findings suggest that the Bethe ansatz only 
accounts for parts of the $S^3$ fluctuation modes, and 
in particular misses out the transverse bosonic and fermionic fluctuations. This may well be not surprising, as the quantum string Bethe ansatz is structurally formulated in a similar way to the SYM Bethe ansatz, and has the same number of degrees of freedom. The corrections proposed in \cite{BT} account for the non-analytic terms (at half-filling), however, it remains unclear, how to systematically find these correction terms, and furthermore, how to incorporate the exponential terms. 

Ideally the present analysis would be done for the stable solution in the $\sltwo$ sector, which was explored and compared to the Bethe ansatz in \cite{Schafer-Nameki:2005tn}, again making use of the infamous zeta-function regularization.  
The semi-classically computed one-loop energy shift in this  
sector \cite{Park:2005ji} is structurally more complicated, however the 
method that we apply here can be expected to also compute the $\sltwo$ case exactly. 
We shall briefly discuss this at the end of the letter. 

The plan of this note is as follows. An outline of the general stratagem in
section 2 is followed by the analysis for the $\sutwo$ sector energy shift in section 3, for which we derive the complete series in $\lambda'$. We conclude with comments on the $\sltwo$ sector.

%\newpage

%%%%%%%%%%%%%%%%%%%%%%%%%%%%%%%%%%%%%%%%%%%%%%%%%%%%%%%%%%%%%%%%%%%%%%%%

\section{The Strategy}

Consider the following often-posed problem: given a sum
$S(\lambda')= \sum_{n\in \mathbb{Z}} f(n, \lambda') $, find the expansion in terms of the parameter $\lambda'$ around zero. Unless the sum converges uniformly, swapping the sum and expansion is not legitimate, wherefore in such an instant one is well-advised to first evaluate the sum and then perform the expansion in $\lambda'$. In order to do so, we shall make use of a nice trick, which relates the sum to a contour integral in the complex plane, and allows to evaluate it by means of complex analytic methods, namely 
\be\label{SumToContour}
2 \pi i \sum_{n\in \mathbb{Z}} f(n, \lambda') = \pi \oint_{\mathcal{C}_r}dz  \cot(\pi
z) f(z, \lambda') \,,
\ee
with the contour $\mathcal{C}_r$ encircling the real axis. 
In case $f(z, \lambda')$ has branch-cuts inside the integration contour, 
the contribution of the integrals around these needs to be subtracted on the LHS.
Subsequently deforming the contour to infinity, one is left with the sum over residues or cut integrals of possible poles and branch-cuts of $f(z, \lambda')$ in the complex plane. 
This method was \eg, applied in the context of light-cone plane-wave string field theory \cite{Lucietti:2003ki, Lucietti:2004wy} and a version of it is known and used in field-theory as the Sommerfeld-Watson transform. The advantage is, that in this way one either obtains a closed expression for the sum, or the limit $\lambda'\rightarrow 0$ can be performed directly on the resulting cut-integrals. 

%%%%%%%%%%%%%%%%%%%%%%%%%%%%%%%%%%%%%%%%%%%%%%%%%%%%%%%%%%%%%%%%%%%%%%%%%

As a sample application consider the simple case of the folded string
\cite{FrolovTseytlinI}, the one-loop energy shift of which is 
\be
\kappa \delta E_{\rm fold} =  -( \sqrt{2} -3) \kappa 
+  {1\over 2} \sum_{n\in\mathbb{Z}} \left(\sqrt{n^2 + 4 \kappa^2}
+ 2 \sqrt{n^2 + 2 \kappa^2} + 5 \sqrt{n^2}  - 8  \sqrt{n^2 + \kappa^2}\right) \,.
\ee
%First note that due to the sole dependence on $n^2$, we have
%\be
%\sum_{n=1}^\infty  f_n = -{1\over 2} f_0 + {1\over 2} \sum_{n=
%  -\infty}^\infty  f_n \,.
%\ee
%In the present case $f_0 = 2 (\sqrt{2 }-3) \kappa$. After rewriting
%the sum of the $f_n$ into a contour integral, we can separate the
%terms into pieces
To evaluate this in a series expansion for $\kappa \rightarrow \infty$, apply (\ref{SumToContour}), then the integrand is made out of terms of the type 
$\sqrt{z^2 + a^2  \kappa^2}$, with
branch-cuts from $\pm i a \kappa $ to $\pm i \infty$. 
Deform the contour to encircle the respective cuts. One can formally do this by introducing a cutoff $\Lambda$ for each cut-integral, which then drops out when summing over the contributions of all the different cuts.
If one is only interested in the non-exponential correction terms, the integral
can be computed by changing to $z=iw \kappa$ and setting $\coth(\pi w
\kappa)$ to one. Performing the integrals yields for each value of $a$
\be
I_a = {1\over 4}\left( 2\Lambda^2 + a^2 \kappa^2 ( 1 + \log (4))  -2 
a^2 \kappa^2 \left( \log (i a \kappa ) + \log (1/\Lambda) \right)\right) \,.
\ee
There is a subtlety for the integral with $a=0$, as the branch-cut in this
case is not along the imaginary axis, but extends from $-\infty$ to
$0$. The integral needs to be computed separately in this case and
yields (in accord with the analytic continuation of the zeta-function)
\be
\tilde{I}_0 = -1/12+ \Lambda^2/ 2\,.
\ee
Adding the terms present in $\delta
E$ together, the divergences and log-terms cancel and we arrive at
\be
\delta E_{\rm fold } 
         = {1\over \kappa} \left(I_2 + 2 I_{\sqrt{2}} + 5\tilde{I}_0 - 8I_1\right)  
         =  -3 \log 2 \ \kappa + (3 - \sqrt{2}) -{5\over 12 \kappa} + O(e^{-\kappa}) \,,
\ee
which indeed agrees with the findings in 
\cite{FrolovTseytlinI, Schafer-Nameki:2005is}. 

\section{The circular string on $S^3 \times \mathbb{R}$}

Let us now apply this method to the circular string in the $\sutwo$ ``subsector''
with two equal spins $J_1 = J_2 =J/2$. The one-loop energy shift was
obtained to be \cite{Frolov:2003tu, Frolov:2004bh, Beisert:2005mq}
\be\label{SUtwoE} 
\delta E^{\sutwo} = \delta E^{(0)} +
\sum_{n=1}^\infty \delta E^{(n)} \,, 
\ee 
where 
\be 
\ba
\delta E^{(0)} 
&= 2 + \sqrt{1-{ 2 k^2\over \mathcal{J}^2 + k^2}} - 3
  \sqrt{1-{k^2 \over \mathcal{J}^2 + k^2}} \cr 
\delta E^{(n)} 
&= 2
  \sqrt{1+ {(n+ \sqrt{n^2 - 4 k^2 })^2 \over 4 (\mathcal{J}^2 + k^2 )}} %\cr 
%& \quad    
 + 2 \sqrt{ 1+ {n^2 - 2 k^2 \over \mathcal{J}^2 + k^2}} + 4
  \sqrt{1 + {n^2 \over \mathcal{J}^2 + k^2}} 
    - 8 \sqrt{1 + {n^2 - k^2 \over \mathcal{J}^2 + k^2}} \label{DeltaE}\,.  
\ea 
\ee 
The various terms in $\delta E^{(n)}$ are in turn: two $S^3$ characteristic frequencies, six transverse bosonic frequencies, and eight fermionic frequencies, which enter with the opposite sign. Furthermore $\lambda' = 1/\mathcal{J}^2$ and we wish to expand this for large $\mathcal{J}$.

To be precise, (\ref{DeltaE}) is the result for even winding. For odd winding the fermions are half-integer moded, the field being antiperiodic \cite{Frolov:2003tu, Frolov:2004bh}. In this case the fermionic fluctuations are to be replaced as follows
\be\label{HalfIntFermi}
\delta E^{\rm fermi} = - 8 \sqrt{1 +
  {n^2 - k^2 \over \mathcal{J}^2 + k^2}}
\rightarrow - 4 \sqrt{1 +
  {(n+ 1/2)^2 - k^2 \over \mathcal{J}^2 + k^2}}- 4 \sqrt{1 +
  {(n- 1/2)^2 - k^2 \over \mathcal{J}^2 + k^2}} \,.
\ee
We shall focus in the main part of the paper on the former and discuss the odd winding in 
appendix A. 
%Denote $\delta E = \sum_{n=1}^\infty \delta E^{(n)}$. 
%The non-zero frequencies were obtained in \cite{Frolov:2003tu} from
%  the roots of certain quartic polynomials. For the bosonic
%  frequencies in the $S^3$ direction (which give the fourth root
%  contribution in (\ref{DeltaE})) as $\sqrt{w_n^{(i)}}$, where $w_n^{(i)}$ are the
%  roots in $w$ of 
%\be
%P_n(w)=  4k^2( n^2 - w )  - ( n^2 - w )^2 +
%4 ( \mathcal{J}^2 + k^2)w) ( -( n^2 - w )^2 - 8k^2w + 4( \mathcal{J}^2
%  + k^2) w  \,.
%\ee
The frequencies appearing in $\delta E$ are real for $2k<1$. We shall
assume this throughout the computation. The result can then be analytically continued 
to other values of $k$ at the end. 

In order to make use of (\ref{SumToContour}), let us first rewrite the 1-loop energy shift as a sum over all integers:
the summands are all dependent only on $n^2$ so that
\be\label{EtoInt}
\ba
\delta E^{\sutwo} \sqrt{\mathcal{J}^2 + k^2} 
&= {1\over 2} 
   \sum_{n=-\infty}^\infty  \left( \omega_n+ \Omega_n  \right) \,.
\ea
\ee
where $\omega_n$ and $\Omega_n$ denotes the $S^3$ fluctuations and transverse/fermionic fluctuations, respectively. Note that the subtraction term from $n=0$ is precisely $\delta E^{(0)} \sqrt{\mathcal{J}^2 + k^2}$. 
Applying (\ref{SumToContour}) yields the contour integral representation
\be\label{EContInt}
\sum_{n\in \mathbb{Z}}  \left( \omega_n + \Omega_n  \right)  
= {1\over 2 i} \left(\oint_{\mathcal{C}_r} - \oint_{\mathcal{C}_2}\right) 
dz \cot(\pi z)  \left(\omega_z  + \Omega_z  \right)  \,.
\ee
The first term is simply the integral around the real axis. In order to understand the second integral, one needs to analyse the cut-structure. Furthermore,
the branch-cuts determine to which contour integrals we can deform the integration along $\mathcal{C}_r$. The branch-cuts of the fermionic and transverse bosonic modes  are
between 
\be\label{TransCuts}
[-i \infty, - i \sqrt{\mathcal{J}^2 \pm k^2}]\, \cup\, [ i \sqrt{\mathcal{J}^2 \pm k^2}, i \infty]\,,\qquad 
[-i \infty, - i \mathcal{J}]\, \cup\, [ i \mathcal{J}, i \infty]\,,
\ee
respectively and thus extend all along the imaginary axis. 
The fluctuations along the $S^3$ are essentially quartic roots, 
and the corresponding two branch-cuts can be aligned in the following non-intersecting fashion
\be
\left[-i \infty, - i {\mathcal{J}^2 \over \sqrt{\mathcal{J}^2 + k^2}}\right]\, \cup\, 
\left[i {\mathcal{J}^2 \over \sqrt{\mathcal{J}^2 + k^2}}, i \infty\right]\,, \qquad
[-2k, 2k] \,.
\ee
The former is of the same type as the imaginary cuts in (\ref{TransCuts}). The latter is a real cut. Due to this branch-cut the corresponding integral around the cut (denoted by $\mathcal{C}_2$) has to be subtracted in (\ref{EContInt}) -- see the LHS of figure 1. 

The first contour around the real axis in  (\ref{EContInt}) can be deformed to encircle the remaining branch-cuts that extend on the imaginary axis, \ie
\be
\sum_{n\in \mathbb{Z}}  \left(\omega_n + \Omega_n  \right)  
= {1\over 2 i} \left(\oint_{\mathcal{C}_1} - \oint_{\mathcal{C}_2}\right) 
  dz \cot(\pi z) \left( \omega_z + \Omega_z  \right)   \,,
\ee
where the deformed contours are depicted in figure 1. Obviously the terms in $\Omega_z$ do not contribute to the integral $\mathcal{C}_2$. In order to evaluate these integrals for large $\mathcal{J}$ we analyse the two contributions separately. In summary we will find the following contributions:

\begin{figure}[t]
\begin{center}
\includegraphics*[width =1.0\textwidth]{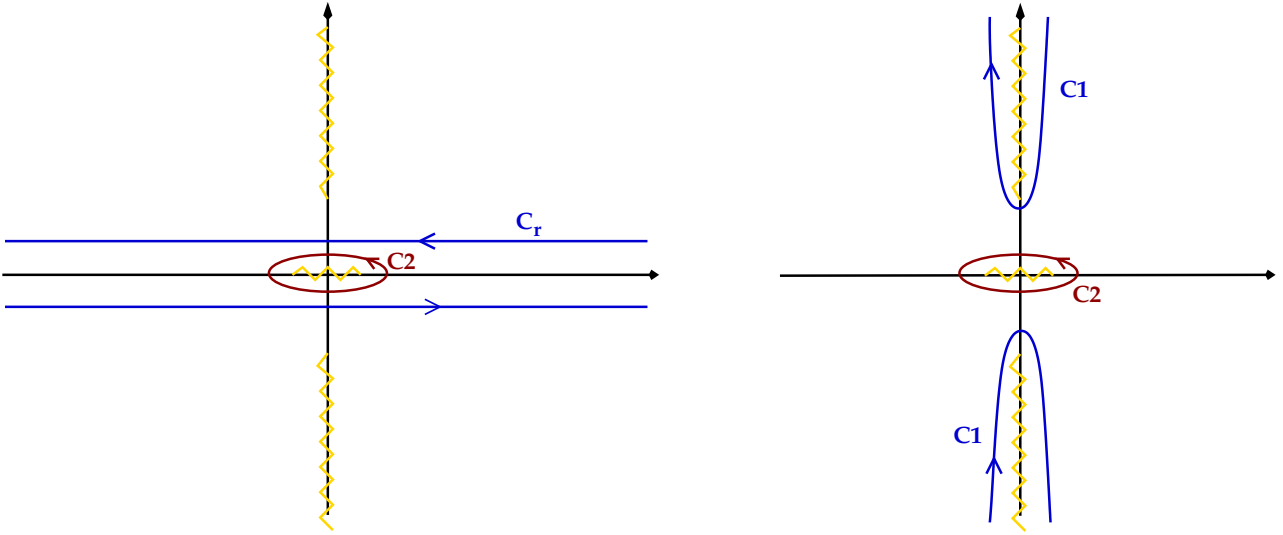}
\end{center}
\caption{Change of integration contours. The branch-cuts are depicted by the yellow
  zig-zag lines.}
\end{figure}

Consider first the integral along $\mathcal{C}_1$. 
As the integration is along the imaginary axis, the cotangent becomes 
a hyperbolic cotangent. Changing variables to $ w= z/\mathcal{J}$, and 
expanding for large $\mathcal{J}$ we can set the cotangent to one and evaluate 
the resulting line integrals. This approximation neglects exponentially small contributions in $\mathcal{J}$ (\cf\  similar discussion in the pp-wave literature \cite{He:2002zu, Lucietti:2003ki, Lucietti:2004wy}). We shall see that the integrals along these cuts yield the non-analytic terms, \ie, of order $1/\mathcal{J}^{2n+1} = {\lambda'}^{n+1/2}$.

The evaluation of the integral along $\mathcal{C}_2$ has to be performed without such an approximation as the cotangent clearly cannot be set to one. We evaluate these by expanding the integrand in $1/\mathcal{J}$ and then integrating up each term with $\cot(\pi z)$ expanded as
\be\label{CotExp}
\cot (x) = {1\over x} + 2 x \sum_{m=1}^\infty {1\over x^2 - \pi^2 m^2} \,.
\ee
This gives precisely the analytic terms as they were computed using zeta-function regularization. 

A remark in view of the analysis of Beisert and Tseytlin \cite{BT} is in order. The split between regular parts and singular parts of the integral approximation and the sum there, 
is precisely the split between the contours $\mathcal{C}_1$ and $\mathcal{C}_2$. 
The above argument makes the there-observed 
agreement between regular and singular parts of the integral and sum, 
respectively, precise. Furthermore our analysis shows that there are exponentially small contributions to the sums, which can be computed as in \cite{He:2002zu, Lucietti:2003ki, Lucietti:2004wy}.

%%%%%%%%

\subsection{Non-analytic terms}

The integrals along $\mathcal{C}_1$ have contributions from all fluctuations, \ie, transverse, fermionic and $S^3$-modes.
Computing the line integrals for the transverse and fermionic modes, with an explicit
(fixed) cut-off $\Lambda$ the integral is
\be
\ba
I_{\rm trans} 
=& \oint_{\mathcal{C}_1^\Lambda}  dw \, \delta E^{\rm trans}(w \mathcal{J}) \cr
=& -8 \log (\mathcal{J}) \mathcal{J}^2+2 \Lambda^2+2 (\mathcal{J}-k) (\mathcal{J}+k) \log (\Lambda) \cr
 &+(\mathcal{J}^2-k^2)\log \left(\mathcal{J}^2-k^2\right)+2 \left(\mathcal{J}^2+k^2\right) \log   \left(\mathcal{J}^2+k^2\right)+(\mathcal{J}^2-k^2)  (1+\log (4))\,.
%
%\frac{1}{2} 
%   \bigg(4 \log (\mathcal{J}^2) \mathcal{J}^2+(1+\log (4)) \mathcal{J}^2-2 
%\Lambda^2 +2 (\mathcal{J}-k) (\mathcal{J}+k) \log (\Lambda) \cr
%&+\left(k^2-\mathcal{J}^2\right) \log\left(\mathcal{J}^2-k^2\right)-2 \left(\mathcal{J}^2+k^2\right) \log \left(\mathcal{J}^2+k^2\right)-k^2 (1+\log (4))\bigg) \,.
\ea
\ee
The regulator dependence will drop out, once we add the contributions from the $S^3$ modes. 

Similarly the case of the frequencies coming from the $S^3$ can be discussed. 
Perform the change of variables suggested by appendix C of 
\cite{Minahan:2005qj}\footnote{ Namely, 
$y= n+ \sqrt{n^2 - 4 k^2}$ for $n>0$ and $y= n- \sqrt{n^2 - 4 k^2}$ for $n<0$. Then the integral from $n=-\Lambda \cdots \Lambda$ extends from 
$y= -\Lambda - \sqrt{\Lambda^2 -4 k^2}\cdots \Lambda + \sqrt{\Lambda -4k^2}$.}. 
The line integral with cut-off $\Lambda$ is straight forwardly computed
\be
\ba\label{SthreeNA}
I_{S^3}
=& \oint_{\mathcal{C}_1^\Lambda} dw \, \delta E^{\rm trans}(w \mathcal{J}) \cr
=& -\log (4) \mathcal{J}^2-\mathcal{J}^2-k^2-2 \Lambda^2
      -2 \left(\mathcal{J}^2-k^2\right) \log (\Lambda)
  +\left(\mathcal{J}^2-k^2\right) \log \left(\mathcal{J}^2+k^2\right)+k^2 \log (4)
\,.
\ea
\ee
Thus
\be
I_{\rm trans} + I_{S^3} =
-8 \log (\mathcal{J}) \mathcal{J}^2+3 \log \left(\mathcal{J}^2+k^2\right) \mathcal{J}^2-2 k^2+\left(\mathcal{J}^2-k^2\right) \log \left(\mathcal{J}^2-k^2\right)+k^2 \log
   \left(\mathcal{J}^2+k^2\right) \,.
\ee
The contribution of these terms to the energy are
\be
\delta E^{\rm na} = {1\over 2 \sqrt{\mathcal{J}^2 + k^2}}
        \left(8 \log (\mathcal{J}) \mathcal{J}^2+2 k^2
        +\left(k^2-\mathcal{J}^2\right) \log \left(\mathcal{J}^2-k^2\right)
        -\left(3 \mathcal{J}^2+k^2\right) \log \left(\mathcal{J}^2+k^2\right)
         \right) \,,
\ee
and have the large $\mathcal{J}$ expansion
\be
\delta E^{\rm na}
 =  -{1 \over 3} k^6 {1\over \mathcal{J}^5}
   +  {1 \over 3} k^8 {1\over \mathcal{J}^7}
   -  {49\over 120} k^{10} {1\over \mathcal{J}^9}
   + O\left({1\over \mathcal{J}^{11}}\right) \,.
\ee
These are the non-analytic terms that were observed to be missed in the naive zeta-function regularization of the energy shift \cite{BT,Schafer-Nameki:2005is, Minahan:2005qj}.

%%%%%%%%

\subsection{Analytic terms}

Finally we are left with the cut-integral around $\mathcal{C}_2$. As remarked earlier, the integral remains along the real axis and thus the cotangent gives non-trivial contributions in the large $\mathcal{J}$ limit. 
Expanding the integrand $\delta E^{(z)}$ for large $\mathcal{J}$ yields 
\be
\ba
\delta E^{(z)}
  =&  \left(k^2-{1\over 2} z^2 + \frac{1}{2} z\sqrt{z^2-4 k^2}\right)
     {1\over \mathcal{J}^2} \cr
  +& \left(\frac{z^4}{8}-\frac{5 k^4}{4}
     -\frac{1}{8} z  \left(2 k^2+z^2\right)\sqrt{z^2-4 k^2} \right) {1\over \mathcal{J}^4}\cr
  +& \left(\frac{1}{16} \left(14 k^6+17 z^2 k^4+2 z^4 k^2-z^6\right)
     + \frac{1}{16} z \left(3 k^4+z^4\right)  \sqrt{z^2-4 k^2} \right) 
      {1\over \mathcal{J}^6} + O\left({1\over \mathcal{J}^8}\right)\,.
\ea
\ee
Integrating each order in $\mathcal{J}$ together 
with $\cot (\pi z)$ in the representation (\ref{CotExp}) yields precisely the zeta-function regularized part, \ie, the analytic in $\lambda'$ terms. Consider first the non-zero mode terms (\ie, the sum part of (\ref{CotExp}))
\be
\ba
{1\over\pi}\oint_{\mathcal{C}_2} dz 
&  \left(  \sum_{n=1}^\infty {2 z \over z^2 -  n^2}\right) 
              \delta E^{(z)} \cr
&=   \sum_{n=1}^\infty \frac{1}{2} \left(2 k^2+n \left(\sqrt{n^2-4 k^2}-n\right)\right)  
                {1\over \mathcal{J}^2 } \cr
&+   \sum_{n=1}^\infty \frac{1}{8} \left(n^4-6 k^4 - n \left(n^2-2 k^2\right)  \sqrt{n^2-4 k^2} 
               \right)  {1\over \mathcal{J}^4 } \cr
&+   \sum_{n=1}^\infty \frac{1}{16} \left(10 k^6-n^2 k^4+2 n^4 k^2-n^6  +n \left(3 k^4+n^4\right) 
     \sqrt{n^2-4 k^2} \right)  {1\over \mathcal{J}^6 } + O\left({1\over \mathcal{J}^8}\right) \,.
\ea
\ee
We should emphasize, that at no time in this computation we made use of zeta-function (or any other) regularization. This result simply provides a rigorous derivation of the energy shift without any unjustified assumptions.

Finally we need to address the zero mode terms. The contribution from $\delta E^{(0)}$ 
in (\ref{SUtwoE}) has already been accounted for in (\ref{EtoInt}). 
So the term that still needs to be accounted for is the $1/z$ ``zero-mode'' term in the expansion of the cotangent (\ref{CotExp}), which has a non-trivial residue at $0$ with the contribution
\be
\pi \hbox{Res}_{z=0} \, \cot (\pi z) \sqrt{4 \mathcal{J}^2 + 4 k^2  + (z + \sqrt{z^2- 4 k^2} )^2} = 2 \mathcal{J} \,.
\ee
This is again analytic in $\lambda'$ and the corresponding term has the expansion
\be
\delta E^{\rm zero} =1 -{k^2 \over 2 }{1\over \mathcal{J}^2} 
                       +{3 k^4 \over 8 }{1\over \mathcal{J}^4} 
                       -{5 k^6 \over 16}{1\over  \mathcal{J}^6}
                    + O\left({1\over \mathcal{J}^8}\right) \,.
\ee
In summary we obtain the rather concise expressions for the energy in an expansion in $1/\mathcal{J}$
\be
\ba
\delta E^{\rm na} &=  {1\over 2 \sqrt{\mathcal{J}^2 + k^2}}
        \left(8 \log (\mathcal{J}) \mathcal{J}^2+2 k^2
        +\left(k^2-\mathcal{J}^2\right) \log \left(\mathcal{J}^2-k^2\right)
        -\left(3 \mathcal{J}^2+k^2\right) \log \left(\mathcal{J}^2+k^2\right)
         \right)\cr
\delta E^{\rm a} &= {1\over \pi }\sum_{i=1}^\infty \left( {1\over \mathcal{J}^{2i}}
                        \sum_{n=1}^\infty \left[ \oint_{\mathcal{C}_2} dz \, {2 z \over z^2 - n^2}
                             \left.\delta E^{(z)} \right|_{\mathcal{J}^{-2i}} \right]
                    \right) \cr
\delta E^{\rm zero} &= {\mathcal{J} \over \sqrt{\mathcal{J}^2 + k^2}} \,.
\ea
\ee
Here, $\left.\delta E^{(z)} \right|_{\mathcal{J}^{-2i}}$ denotes the coefficient of $1/\mathcal{J}^{2i}$ in the expansion of $\delta E^{(z)}$. Recall also that in the expression for the non-analytic terms $\delta E^{\rm na}$ is exact up to exponential corrections $O\left(e^{-\mathcal{J}}\right)$. $\delta E^{\rm a}+ \delta E^{\rm zero}$ reproduce the terms that one obtains naively from zeta-function regularization.

%%%%%%%%%%%%%%%%%%%%%%%%%%%%%

\subsection{Exponential corrections}

The exponential corrections have so far been neglected in the contour integral along $\mathcal{C}_1$ by setting $\cot (\pi \mathcal{J}z)$ for imaginary $z$ and large $\mathcal{J}$ to one. 
Here, we wish to determine an exact formula for them.
The strategy is to differentiate $\delta E^{(n)}$ twice with respect to $\mathcal{J}$. 
By this procedure we can treat each frequency separately, as each separate sum converges, although we loose information about the polynomial dependence on 
$1/\mathcal{J}$. However as we have explicit expressions for these to all orders already we can safely ignore this issue. 
For the transverse and fermionic fluctuations the relevant terms after acting with
$\mathfrak{O}=\left({1\over \mathcal{J}}{\partial \over \partial \mathcal{J}}\right)^2$ is, for $a=a(\mathcal{J})$
\be
S(a) = - \sum_{n=1}^\infty {1\over (a^2 + n^2)^{3/2}} \,.
\ee
Applying (\ref{SumToContour}) to this sum yields after integration by parts 
\be
S(a) = -{1\over a^2} + {1\over 2 a^3} - {\pi \over a}  
         \int_1^\infty dz {z \over \sqrt{z^2-1}} {1\over \sinh^2 (\pi a z )} \,.
\ee
The combined expressions for the transverse modes, integrated up again is
\be\label{ExpTrans}
\ba
& \delta E^{\text{trans}}|_{\text{exp}} \cr
& = {1\over {\sqrt{\mathcal{J}^2 + k^2}}} 
  \left. \mathfrak{O}^{-1}  
    \left( 8 S(\mathcal{J}) - 4 S(\sqrt{\mathcal{J}^2 + k^2}) 
        - 2 S(\sqrt{\mathcal{J}^2 - k^2}) \right) \right|_{\text{exp}}\cr
& ={1\over 2 \pi^2 \sqrt{\mathcal{J}^2+k^2} }
 \int_{1}^\infty  {dz \over  z^2 \sqrt{z^2-1}} \times\cr
 & \qquad \qquad \qquad  \times \sum_{l=0, k, i k} \alpha_l
    \left[\left(2 \pi z  \sqrt{\mathcal{J}^2+ l^2}  \log 
       \left(1-e^{-2 \sqrt{\mathcal{J}^2+l^2} \pi  z}\right)
    -\text{Li}_2\left(e^{-2 \sqrt{\mathcal{J}^2+l^2} \pi z}\right)\right)\right] \,.
\ea
\ee
where $\alpha_0= 8, \alpha_k =-4, \alpha_{i k} = -2 $
and the polylogs $\text{Li}_n = \sum_{m=1}^\infty {z^m \over m^n}$.
Since the dependence on $\mathcal{J}$ is now only in the prefactor and the (poly)log-terms, the exponential corrects are obtained by expanding the log in a power series in $e^{-\mathcal{J}}$. 

The remaining term from the $S^3$-fluctuations are 
\be\label{ExpSphere}
\ba
&\delta E^{S^3}|_{\text{exp}}
 =\left. - 4 \mathfrak{O}^{-1}\sqrt{2} \int_{\mathcal{C}_2} 
   dz {1\over (2 \mathcal{J}^2 + z(z+ \sqrt{z^2 - 4 k^2}))^{3/2}} \cot(\pi z) 
          \right|_{\text{exp}}\cr
%&=  -\mathfrak{O}^{-1} {2 \over (\mathcal{J}^2 + k^2) }
%         \int_{1}^\infty dy 
%        \left({k^2 \over (\mathcal{J}^2+k^2) y^2}+1 \right){1\over (y^2-1)^{3/2}} 
%    \coth\left(\sqrt{\mathcal{J}^2 + k^2 } \pi y 
%                   -{k^2 \over \sqrt{\mathcal{J}^2 + k^2}}{1\over y} \right) \cr
&=  {1\over \pi \sqrt{\mathcal{J}^2 + k^2}}
   \int_{1}^\infty dy {y\over  \sqrt{y^2-1}} \int^\mathcal{J} d\mathcal{J} 
    {\mathcal{J} \over \left(\mathcal{J}^2+k^2\right)^{3/2}} 
   \log \left(1-e^{-\frac{2 \pi  
    \left(\left(y^2-1\right) k^2+\mathcal{J}^2 y^2\right)}{\sqrt{\mathcal{J}^2+k^2}   y}}\right) \,.
\ea
\ee
The combined expressions (\ref{ExpTrans}) and (\ref{ExpSphere})  are the exponentially suppressed terms in the string energy shift. 

%%%%%%%%%%%%%%%%%%%%%%%%%%%%%%%%%%%%%%%%%%%%%%%%%%%%%%%%%%%%%%%%%%%%%%%%%%%

\subsection{Comments on the $AdS_3 \times S^1$ sector}

From the foregoing analysis we can learn various points about the spinning strings on $AdS_3 \times S^1$.
This case is of interest, as on does not require the analytic continuation in the winding number $k$ that we had to make use of for the $S^3\times \mathbb{R}$ case. In particular the solutions in this sector are stable for all values of the winding numbers $k$ and $m$ (we refer the reader to \cite{Park:2005ji, Schafer-Nameki:2005tn} for the notation used). The $AdS_3$ fluctuations are the obstruction to exactly evaluating the sum in this case, and they are given by
$\sum_{I=1}^4 \epsilon_{n, I}\omega_{n}^{(I)}$, where $\omega_n^{(I)}$ are roots of a quartic polynomial $P_n(\omega)$, and $\epsilon_{n,I}$ are signs. As a function of $n$ these are complicated non-analytic functions, and unfortunately we have nothing much to say about these.  
However 
the structure of the transverse and fermionic fluctuations is similar to the ones in 
(\ref{DeltaE}). In particular, these contribute only through branch-cuts of the type $\mathcal{C}_1$ and can be treated in an identical fashion to section 3.1. One finds that these again contribute with odd powers of $1/\mathcal{J}$ and exponential terms, which when compared to \cite{Schafer-Nameki:2005tn} are again yet to be included into the Bethe ansatz.

%%%%%%%%%%%%%%%%%%%%%%%%%%%%%%%%%%%%%%%%%%%%%%%%%%%%%%%%%%%%%%%%%%%%%%%%%%%

\subsection*{Acknowledgments}

I thank S.~Frolov, A.~Tseytlin, M.~Zamaklar and K.~Zarembo 
for discussions and for comments on the draft.
This work was partially supported by
the DFG, DAAD, and European RTN Program MRTN-CT-2004-503369.

%%%%%%%%%%%%%%%%%%%%%%%%%%%%%%%%%%%%%%%%%%%%%%%%%%%%%%%%%%%%%%%%%%%%%%%%%%%%

\setcounter{section}{0}
\appendix{Half-integral fermion frequencies}

In this appendix we discuss the energy shift for the circular string in the $\sutwo$ sector with the half-integer moded fermion frequencies (\ref{HalfIntFermi}), which arise for odd winding number. We shall confine our analysis to the non-analytic terms. Needless to say, the analytic terms are unchanged. 

The main change to note is that the branch-cuts for the frequencies $\sqrt{(n\pm 1/2)^2 + \mathcal{J}^2}$ are located from 
\be
[1/2 - i \infty , 1/2 - i \mathcal{J}] \cup [1/2 + i \mathcal{J}, 1/2 + i \infty]\,,\qquad 
[-1/2 - i \infty , -1/2 - i \mathcal{J}] \cup [-1/2 + i \mathcal{J}, -1/2 + i \infty]
\ee
Together with the remaining cut-integrals for the other transverse modes, these contribute
\be
\ba
I^{\rm trans} =&
\Lambda^2
- (\mathcal{J}^2-k^2) \log (\Lambda)
+\frac{1}{2} \bigg(
8 \log (\mathcal{J}) \mathcal{J}^2+\left(k^2-\mathcal{J}^2\right) \log \left(\mathcal{J}^2-k^2\right) \cr 
&\qquad \qquad -2 \left(\mathcal{J}^2+k^2\right) \log \left(\mathcal{J}^2+k^2\right)+\left(k^2-\mathcal{J}^2\right) (1+\log (4))\bigg)
\ea
\ee
$\Lambda$ is again the cut-off. The $S^3$-fluctuation frequencies are unchanged (\ref{SthreeNA}) and joining these we arrive at
\be
\delta E^{\rm na} ={1\over 2 \sqrt{\mathcal{J}^2 + k^2}} 
 \left(8 \log (\mathcal{J}) \mathcal{J}^2+2 k^2+\left(k^2-\mathcal{J}^2\right) \log \left(\mathcal{J}^2-k^2\right)-\left(3 \mathcal{J}^2+k^2\right) \log
   \left(\mathcal{J}^2+k^2\right)\right)
\ee
This agrees with the correction for the integer-moded fermion case in the main part of the paper.

%%%%%%%%%%%%%%%%%%%%%%%%%%%%%%%%%%%%%%%%%%%%%%%%%%%%%%%%%%%%%%%%%%%%%%%%%%%%

%\newpage

%%%%%%%%%%%%%%%%%%%%%%%%%%%%%%%%%%%%%%%%%%%%%%%%%%%%%%%%%%%%%%%%%%%%%%%%%%%%

\bibliographystyle{JHEP} \renewcommand{\refname}{References}
\addcontentsline{toc}{section}{Bibliography} 
%\bibliography{main}

\providecommand{\href}[2]{#2}\begingroup\raggedright\endgroup

\end{document}